\documentclass[conference]{IEEEtran}
\IEEEoverridecommandlockouts

\usepackage{makecell}
\usepackage{multirow}
\usepackage{cite}
\usepackage{amsmath,amssymb,amsfonts}
\usepackage{algorithmic}
\usepackage{amsmath}
\usepackage{amssymb}
\usepackage{pifont}
\usepackage[normalem]{ulem}
\usepackage{booktabs}

\newcommand{\tabincell}[2]{%
  \begin{tabular}{@{}#1@{}}#2\end{tabular}%
}
\newcommand{\cmark}{\ding{51}}
\newcommand{\xmark}{\ding{55}}

\usepackage{graphicx}
\usepackage{textcomp}
\usepackage{xcolor}
\def\BibTeX{{\rm B\kern-.05em{\sc i\kern-.025em b}\kern-.08em
    T\kern-.1667em\lower.7ex\hbox{E}\kern-.125emX}}
\begin{document}

\title{Differentiable and Severity-invariant Discrete Tokens for Dysarthric Speech Recognition
}

\author{
\IEEEauthorblockN{
Huimeng Wang\IEEEauthorrefmark{1},
Xurong Xie\IEEEauthorrefmark{2},
Mengzhe Geng\IEEEauthorrefmark{3},
Haoning Xu\IEEEauthorrefmark{1},
Jiajun Deng\IEEEauthorrefmark{1},\\
Youjun Chen\IEEEauthorrefmark{1},
Chengxi Deng\IEEEauthorrefmark{1},
Xunying Liu\IEEEauthorrefmark{1}
}
\IEEEauthorblockA{
\IEEEauthorrefmark{1} The Chinese University of Hong Kong,
Hong Kong SAR, China\\
\IEEEauthorrefmark{2} Institute of Software, Chinese Academy of Sciences,
Beijing, China\\
\IEEEauthorrefmark{3} National Research Council Canada,
Canada
}
}

\maketitle

\begin{abstract}
This paper proposes novel differentiable and severity-invariant (DSI) discrete token approaches that are not only tightly integrated with downstream dysarthric speech recognition tasks, but also minimise discrete token diversity across speech impairment severity groups.
Experiments conducted on the UASpeech and TORGO corpora suggest that Conformer models trained using the DSI tokens outperform the comparable baseline HuBERT discrete/continuous features by statistically significant WER reductions of 2.22\%/0.78\% absolute (9.14\%/3.41\% relative) and 1.78\%/1.06\% absolute (18.43\%/11.86\% relative) on the two tasks, respectively. After system combination, the lowest WERs of 18.90\% and 6.38\% were obtained on UASpeech and TORGO.
Phoneme-specific T-SNE visualizations show that severity-invariant regularization reduces severity-dependent variation by producing greater overlap and less distinct boundaries among severity-group distributions.
\end{abstract}


\section{Introduction}
Despite the rapid advancements in automatic speech recognition (ASR) technologies targeting normal speech, accurate recognition of pathological voice, for instance, dysarthric speech, remains to date a highly challenging task \cite{xiong20source,mengzhetaslp,shujie23taslp,wang2023benefits,huimeng2024icassp,huimeng2025icassp,hu25d_interspeech,xiong25_interspeech,hsieh25_interspeech} due to: 
\textbf{a)} the scarcity of such data; \textbf{b)} its large mismatch against the normal speech; and \textbf{c)} large speaker-level diversity.
The physical disabilities and mobility issues associated with impaired speakers exacerbate the challenges of collecting large quantities of dysarthric speech for ASR system development.
To this end, a series of recent research has focused on migrating knowledge from normal speech ASR systems to tackle dysarthric speech recognition tasks \cite{mengzhetaslp,shujie23taslp,huimeng2024icassp,huimeng2025icassp,hu25d_interspeech,xiong25_interspeech,hsieh25_interspeech,aboeitta25_interspeech}.

In recent years, self-supervised learning (SSL) based speech foundation models \cite{baevski2020wav2vec,hsu2021hubert,chen2022wavlm} pre-trained on massive unlabeled data have emerged as a powerful paradigm for multiple downstream speech processing tasks.
Neural speech representations from these models are highly adaptable across different task domains.
In particular, compact discrete token features have been successfully applied to a wide variety of downstream tasks including automatic speech recognition \cite{chang23b_interspeech,chang2024exploring,yang2024towards,onda25c_interspeech, cui25_interspeech, sukhadia24_interspeech}, text-to-speech synthesis \cite{yang2024towards,du22b_interspeech,unicats}, accent normalization \cite{bai25_interspeech}, and voice conversion \cite{wang25ba_interspeech}.

Despite consistently underperforming their continuous counterparts across different tasks, discrete tokens offer several advantages that motivate their adoption in dysarthric ASR.
The discrete nature of speech tokens enables efficient sequence compression that effectively mitigates the computational overhead caused by dense continuous representations.
They also provide a natural interface for unified text-speech modeling in SpeechLLM-based dysarthric recognition systems.
Besides, discrete tokens inherently reduce storage and transmission costs for cloud-based deployment, while also supporting on-device inference for privacy-sensitive pathological speech recognition under limited computational resources.

Efforts to apply discrete tokens for dysarthric speech recognition are confronted with the aforementioned severe performance degradation.
This issue is specifically driven by underlying limitations:
\textbf{1) Misaligned training objectives} between discrete token extraction and ASR backend cause task-relevant information loss \cite{yeh2024slt}.
Although end-to-end (E2E) optimized differentiable K-means is applied to mitigate the training inconsistency, a notable performance gap persists on healthy, non-elderly speech \cite{onda25c_interspeech}.
\textbf{2) The large pathology-induced speaker heterogeneity} in dysarthric speech further exacerbates the performance gap.
Common sources of variation, such as accent or gender, when further compounded with the speech and language pathology severity, create large diversity among dysarthric speakers \cite{mengistu2011adapting}.
This diversity presents a challenge that the E2E optimized K-means tailored for typical speech domains is ill-equipped to handle.

To this end, this paper proposes novel differentiable and severity-invariant (DSI) discrete tokens for dysarthric speech recognition.
These tokens are extracted through three key approaches: 
\textbf{1) iterative discrete pseudo-label update} for token refinement;
\textbf{2) end-to-end optimization} that tightly integrates differentiable token extraction with the Conformer \cite{gulati20_interspeech} backend to learn ASR-oriented discrete tokens;
and \textbf{3) severity-invariant regularization (SIR)} that minimizes the differences between utterance-level discrete tokens that are computed across speech impairment severity groups.

Experiments conducted on the benchmark UASpeech \cite{uaspeech2008} and TORGO \cite{torgo} dysarthric corpora suggest that Conformer models trained using DSI tokens outperform the comparable baseline HuBERT discrete/continuous features by statistically significant WER reductions of 2.22\%/0.78\% absolute (9.14\%/3.41\% relative) and 1.78\%/1.06\% absolute (18.43\%/11.86\% relative) on the two tasks, respectively. 
After system combination, the lowest WERs of 18.90\% and 6.38\% are obtained on UASpeech and TORGO.
T-SNE analysis further supports the effectiveness of SIR.
Experiments on Qwen-based SpeechLLM systems confirm that DSI tokens generalize across backends, achieving consistent WER reductions on both tasks.

Our main contributions are summarized as follows:
\newline
\noindent 1) This paper proposes novel differentiable and severity-invariant discrete token features for dysarthric speech recognition.
A set of key approaches is integrated to learn both ASR-oriented and pathology severity-invariant representations during token extraction.
In contrast, prior research predominantly focused on utilizing continuous SSL speech representations \cite{shujie23taslp, wang2023benefits, hsieh25_interspeech, hernandez22_interspeech, 10081405} to improve dysarthric speech recognition performance.
Very limited prior research \cite{huimeng2025icassp} exploring discrete token based dysarthric speech recognition ignores both training inconsistency and significant heterogeneity inherent in dysarthric speech, as considered in this paper.
Although E2E optimization \cite{onda25c_interspeech} has been proposed for normal speech to address inconsistencies, the heterogeneity posed by speech pathology severity has been largely overlooked.

\noindent 2) This paper proposes the severity-invariant regularization to reduce severity-dependent variation in discrete token representations. Beyond ASR performance gains, phoneme-specific T-SNE visualizations show that the regularized representations lead to greater overlap and less distinct boundaries among severity-group distributions.

\noindent 3) Extensive experiments on UASpeech and TORGO demonstrate that the proposed discrete tokens consistently improve dysarthric ASR performance across different codebook sizes and ASR backends, achieving competitive results over both continuous SSL representations and conventional discrete token baselines.

\begin{figure*}[!t]
    \centering    \includegraphics[width=0.75\linewidth]{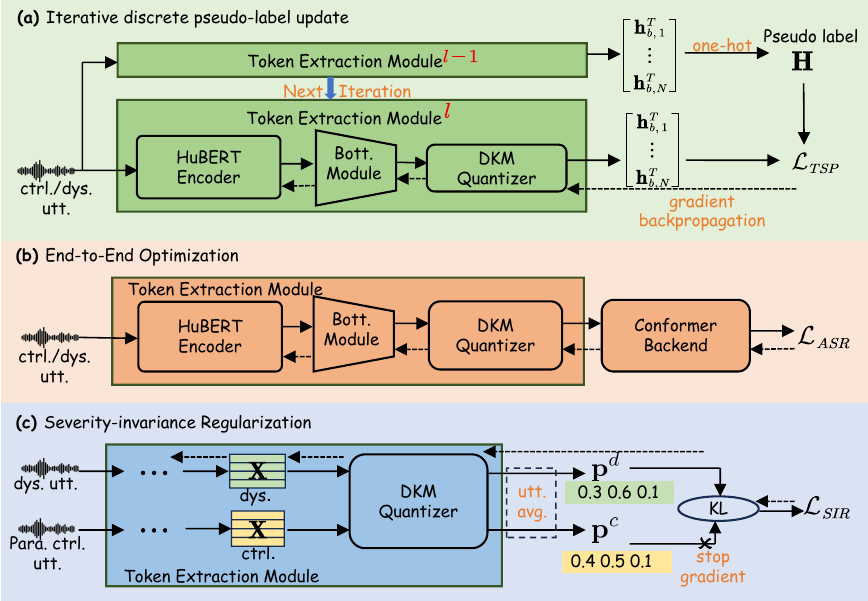}
    \vspace{-2mm}
\caption{
Overview of the proposed differentiable and severity-invariant (DSI) discrete token learning framework.
\textbf{(a)} \textbf{Iterative pseudo-label update} for progressive token refinement.
\textbf{(b)} \textbf{End-to-end optimization} between the differentiable token extraction module and the Conformer ASR backend for learning ASR-oriented discrete tokens.
\textbf{(c)} Additional \textbf{severity-invariant regularization} applied during E2E optimization, which minimizes the divergence between utterance-level token distributions of content-parallel dysarthric and healthy speech.
}
\label{fig:1}
\vspace{-4mm}
\end{figure*}

\section{Foundation models and Discrete Tokens}
\label{sec:foundation-models-and-discrete-tokens}
\subsection{HuBERT}

\noindent\textbf{Model architecture:} HuBERT comprises: 1) a CNN encoder processing speech $\mathbf{O}$ into continuous representations $z_t \in \mathcal{Z}$, with a $25$ms receptive field and $20$ms stride;
2) a Transformer and projection layer mapping randomly masked $z_t$ to context representations $c_t \in C$; and 3) a quantization module producing speech units $q_t \in \mathcal{Q}$ as training pseudo-labels.

\noindent\textbf{Model pre-training:} of HuBERT iteratively alternates between two steps: a) generating frame-level training pseudo-labels via offline clustering (first on MFCC, then on its intermediate representations) and b) optimizing a BERT-like masked prediction objective using these pseudo-labels.

\noindent\textbf{Model fine-tuning:} During fine-tuning, the projection layer is substituted with a randomly initialized softmax layer. 
The model parameters are then optimized on labeled speech data using the Connectionist Temporal Classification (CTC) loss $\mathcal{L}_{\mathrm{CTC}}$, with the CNN-based feature encoder kept frozen.

\subsection{Discrete Token Extraction}
Discrete token extraction involves two steps: 1) extracting continuous speech representations, and 2) quantizing these features into discrete tokens.

\noindent\textbf{Continuous speech representations} are extracted from speech via a cross-domain fine-tuned HuBERT, modified by inserting a bottleneck module between its Transformer and CTC networks~\cite{shujie23taslp}.
This module comprises four interleaving layers:
1) a 1D transposed CNN layer adjusting the stride from 20 ms to 10 ms;
2) a fully connected (FC) block (linear, ReLU, dropout) reducing the representation dimension from 1024 to 256;
3) a second CNN layer reverting the stride back to 20 ms; and 
4) a final FC block restoring the dimension to 1024.
The first FC block's outputs serve as the final continuous speech representations.

\noindent\textbf{Discrete tokens} are derived from continuous speech representations using a K-means codebook.

\section{Differentiable Discrete Tokens}
\label{sec:diff-severity}

\subsection{Differentiable K-means}
\label{subsection:differentiable-kmeans}
Differentiable K-means (DKM) trains a learnable codebook $\mathbf{C} \in \mathbb{R}^{K\times D},$ by minimizing the Euclidean distance between continuous representations $\mathbf{X} \in\mathbb{R}^{B \times N \times D}$, and their soft-assigned centroids.
Here, $K$, $D$, $B$ and $N$ denote the codebook size, feature dimension, batch size, and sequence length.

\noindent\textbf{Soft assignment}: For each frame $\mathbf{x}_{b,n}$, DKM computes a soft assignment probability for each of the $K$ centroids $\mathbf{c}_k$ in the codebook.
As proposed in \cite{gao2020deep}, this probability is calculated using the negative Euclidean-distance-based softmax function:
\begin{equation}
P(k|\mathbf{x}_{b,n}, \mathbf{C}) = \frac{\exp\left(-\sigma^2 \lVert \mathbf{x}_{b,n} - \mathbf{c}_k \rVert^2_2\right)}{\sum_{j=1}^{K} \exp\left(-\sigma^2 \lVert \mathbf{x}_{b,n} - \mathbf{c}_j \rVert^2_2\right)}
\end{equation}

\noindent where the hyperparameter $\sigma$ is set to 1.0. 
We then use the Gumbel-Softmax trick \cite{jang2016categorical} to sample a relaxed assignment vector $\mathbf{h}_{b,n} \in \mathbb{R}^K$. Its $k$-th element is computed as:
\begin{equation}
h_{b,n}^k = \frac{\exp \left( \big(\log P(k \mid \mathbf{x}_{b,n}, \mathbf{C}) + G_k\big) / \tau \right)}{\sum_{j=1}^K \exp \left( \big(\log P(j \mid \mathbf{x}_{b,n}, \mathbf{C}) + G_j\big) / \tau \right)} \label{eq:3}
\end{equation}
\noindent where $G_k$ are i.i.d. samples drawn from the standard Gumbel distribution, and $\tau$ is the temperature.

\noindent\textbf{Euclidean-distance-based K-means loss} is defined as the distance between the features $\mathbf{X}$ and their selected centroids:
\begin{equation}
    \mathcal{L}_{\mathrm{KM}} = \sum_{b=1}^{B}\sum_{n=1}^{N} \lVert \mathbf{x}_{b,n} - \mathbf{C}^T\tilde{\mathbf{h}}_{b,n} \rVert^2_2
\end{equation}
where $\tilde{\mathbf{h}}_{b,n}$ is the one-hot version of $\mathbf{h}_{b,n}$ obtained via a straight-through $\operatorname{argmax}$ operation.
The codebook $\mathbf{C}$ is subsequently optimized by minimizing the loss function $\mathcal{L}_{\mathrm{KM}}$.
Notably, the loss $\mathcal{L}_{\mathrm{KM}}$ is used only for the initial DKM training stage.
Subsequent pseudo-label update and E2E optimization are performed using training objectives described in Sec. \ref{tokenrefinement} and Sec. \ref{e2eoptimization}.

\subsection{Iterative Discrete Pseudo-label Update}\label{tokenrefinement}
The iterative discrete pseudo-label update illustrated in Fig.~\ref{fig:1}(a) progressively improves the quality of discrete tokens through an alternating optimization procedure. Starting from an initial token set $\mathbf{H}^{1}$, each iteration consists of (1) a token-supervised quantizer update, which optimizes the HuBERT encoder and DKM quantizer using the current token labels $\mathbf{H}^{l}$, and (2) a discrete token update, which generates a refined token set $\mathbf{H}^{l+1}$ using the updated models.

\noindent\textbf{Token-supervised quantizer update}: 
At iteration $l$, the current discrete token set $\mathbf{H}^{l}$ is treated as pseudo-label supervision.
Given an input speech utterance $\mathbf{O}$, the HuBERT encoder $\text{Enc}_{\mathrm{HB}}$ first extracts continuous speech representations, which are subsequently mapped into token posterior distributions by the DKM quantizer $\text{Q}_{\mathrm{DKM}}$ as $\mathbf{P}^{l} =
  \text{Q}_{\mathrm{DKM}}(\text{Enc}_{\mathrm{HB}}(\mathbf{O}))$.
The HuBERT encoder and DKM quantizer are then jointly optimized to predict the pseudo-label tokens $\mathbf{H}^{l}$ by minimizing the token prediction loss:
\begin{equation}
\mathcal{L}_{\mathrm{TSP}}
=
-\sum_{n=1}^{N}
\sum_{k=1}^{K}
H^{l}_{n,k}
\log P^{l}_{n,k}
\end{equation}
During this stage, the CNN feature encoder of HuBERT are frozen, while the remaining HuBERT parameters and the DKM codebook are jointly updated.
The initial pseudo-labels $\mathbf{H}^{1}$ are obtained by clustering the continuous HuBERT representations using DKM.

\noindent\textbf{Discrete token update}: 
After the token-supervised update at iteration $l$, the updated HuBERT encoder and DKM quantizer are used to re-tokenize the entire training corpus.
Specifically, continuous speech representations are first extracted by $\text{Enc}_{\mathrm{HB}}$ and subsequently quantized by $\text{Q}_{\mathrm{DKM}}$.
For each frame, a hard token assignment is obtained via the $\operatorname{argmax}$ operation over the quantizer outputs, producing a refined token set $\mathbf{H}^{l+1}$:
\begin{equation}
\mathbf{H}^{l+1}
=
\operatorname{argmax}
\left(
\text{Q}_{\mathrm{DKM}}
(
\text{Enc}_{\mathrm{HB}}(\mathbf{O})
)
\right)
\end{equation}
The refined tokens $\mathbf{H}^{l+1}$ can then serve as pseudo-label supervision for the next iteration, or be directly used as downstream ASR input features.

\subsection{End-to-End Optimization between Discrete Token Extraction and Conformer ASR}\label{e2eoptimization}

To derive ASR-oriented discrete tokens, we further perform end-to-end optimization of the discrete token extraction module together with the downstream Conformer ASR backend, as illustrated in Fig.~\ref{fig:1}(b).
Specifically, the HuBERT encoder $\text{Enc}_{\mathrm{HB}}$, DKM quantizer $\text{Q}_{\mathrm{DKM}}$, and Conformer-based ASR model $\text{ASR}_{\mathrm{CFM}}$ are jointly optimized using ASR supervision.

Given a speech utterance $\mathbf{O}$, continuous speech representations are first extracted by $\text{Enc}_{\mathrm{HB}}$ and subsequently converted into discrete token representations through $\text{Q}_{\mathrm{DKM}}$.
The resulting token sequence is then fed into the downstream Conformer ASR backend to predict the corresponding transcript.
Given the ground-truth transcript $\mathbf{Y}$, the entire pipeline is optimized using the multi-task ASR objective
$\mathcal{L}_{\mathrm{ASR}}$\footnote{$\mathcal{L}_{\mathrm{ASR}} = 0.3\mathcal{L}_{\mathrm{CTC}}+0.7\mathcal{L}_{\mathrm{Att.}}$, where $\mathcal{L}_{\mathrm{Att.}}$ is attention loss.}:
\begin{equation}
\mathcal{L}_{\mathrm{E2E}}
=
\mathcal{L}_{\mathrm{ASR}}
\Big(
\mathbf{Y};
\text{ASR}_{\mathrm{CFM}}
(
\text{Q}_{\mathrm{DKM}}
(
\text{Enc}_{\mathrm{HB}}
(\mathbf{O})
)
)
\Big)
\end{equation}
The ASR loss is back-propagated through the entire pipeline, allowing the HuBERT encoder, DKM quantizer, and Conformer backend to be jointly optimized toward improved recognition performance.
The HuBERT encoder and DKM quantizer are initialized using the parameters obtained from the final iteration of pseudo-label refinement, while the Conformer ASR backend is randomly initialized.
Following the standard HuBERT fine-tuning setup, the CNN feature extractor remains frozen throughout training.

\section{Severity-invariant Discrete Tokens}
\label{sec:severity-invariant-token}
As shown in Fig.~\ref{fig:1}(c), the E2E optimization framework is further augmented with a severity-invariant regularization term $\mathcal{L}_{\mathrm{SIR}}$ to derive severity-invariant discrete tokens:
\begin{equation}\label{eq:7}
    \mathcal{L}_{\mathrm{DSI}} = \mathcal{L}_{\mathrm{E2E}} + \alpha \cdot \mathcal{L}_{\mathrm{SIR}}
\end{equation}
\noindent where $\alpha$ is an empirically set hyperparameter.

The \textbf{severity-invariant regularization (SIR)} term $\mathcal{L}_{\mathrm{SIR}}$ in Equation (\ref{eq:7}), aims to minimize the discrepancy between the token distributions of dysarthric speech and its healthy counterparts.
For each dysarthric utterance $\mathbf{o}_b^d$ in a batch, we compute the KL divergence between its utterance-level averaged token distribution $    \mathbf{p}_{b}^{d} = \frac{1}{N_{b}^{d}} \sum_{n=1}^{N_{b}^{d}} \mathbf{h}_{b,n}^d$ and that of a content-parallel healthy utterance $\mathbf{o}_b^c$ from a randomly selected control speaker.
Here, $N_b^d$ denotes the length of the encoded continuous representation sequence, and $\mathbf{h}_{b,n}^d$ is the soft assignment vector from Eq.~(\ref{eq:3}).
The averaged distribution for the parallel healthy utterance $\mathbf{p}^c_b$, is computed similarly during training. 
The final severity-invariant regularization $\mathcal{L}_{\mathrm{SIR}}$ term is given by:
\begin{equation}
    \mathcal{L}_{\mathrm{SIR}} = \frac{1}{B} \sum_{b=1}^B D_{\mathrm{KL}} \left( \operatorname{sg}[\mathbf{p}_{b}^{c}] \, \parallel \, \mathbf{p}_{b}^{d} \right)
\end{equation}
\noindent where $B$ is the batch size and $\operatorname{sg}[\cdot]$ denotes the stop-gradient operation applied to the healthy control distribution.
By minimizing the divergence between dysarthric and healthy token distributions, the proposed regularization encourages the learned discrete representations to capture linguistic content while reducing severity-dependent variations.

\section{Experiments}
\label{experiments}
\subsection{Task Description}
\label{subsection:task}
The English UASpeech dysarthric corpus, an isolated word recognition task,  contains 103 hours of speech from 16 dysarthric speakers and 13 healthy control speakers.
For every speaker, the data is partitioned into three blocks (B1, B2, and B3), each containing a shared set of 155 common words and a unique set of 100 uncommon words.
B1 and B3 from all speakers, alongside B2 from controls, constitute the training set, while B2 from dysarthric speakers forms the test set.
After HTK-based silence stripping \cite{young2002htk} and speed perturbation augmentation \cite{huimeng2024icassp},  the final training and test sets comprise 173 hours (538,292 utt.) and 8.7 hours (26,520 utt.), respectively.

The TORGO dysarthric corpus contains 13.5 hours of speech (5.8h sentences, 7.7h single words) from 8 dysarthric and 7 control speakers, encompassing 1573 distinct words.
Following a partition strategy similar to UASpeech, the training set incorporates all control data and two-thirds of the dysarthric data.
The remaining one-third of dysarthric speech is used as the test set.
After silence stripping and data augmentation \cite{geng2020investigation}, the final training and test sets contain 34.1 hours (61,813 utt.) and 1.0 hour (1,892 utt.), respectively.

\subsection{Experiment Setup}
\label{subsection:setup}
\noindent\textbf{Model Configurations}: 
HuBERT-Large\footnote{https://huggingface.co/facebook/hubert-large-ls960-ft} is used as the SSL backbone for extracting both continuous SSL features and discrete token features in this paper.
The E2E Conformer systems are implemented using the ESPnet toolkit \cite{watanabe18_interspeech}. The hybrid LF-MMI factored TDNN systems \cite{povey16_interspeech} are trained using the Kaldi toolkit, following the standard chain recipe without i-vectors.
The SpeechLLM-based ASR systems are built using the Qwen2.5-0.5B-Instruct model\footnote{https://huggingface.co/Qwen/Qwen2.5-0.5B-Instruct}.

\noindent\textbf{Hyper-parameters}: In the DKM quantizer, the codebook size $K$ is evaluated at $\{100,500\}$, and the temperature $\tau$ in Eq.~(\ref{eq:3}) is fixed at 0.8. The regularization weight $\alpha$ in Eq.~(\ref{eq:7}) is tuned for each experimental setting.

\noindent\textbf{Baselines}: We compare the DSI tokens against continuous SSL representations and conventional K-means-based discrete tokens.
To ensure a fair comparison, all feature representations are evaluated using the same Conformer ASR architecture.

\noindent\textbf{T-SNE visualization}:
For t-SNE visualization, frame-level phoneme labels are obtained by forced alignment~\cite{young2002htk}. For each target phoneme, the corresponding feature frames are averaged within each utterance to form an utterance-level phoneme-specific representation. Representations with and without severity-invariant regularization are jointly projected into the same t-SNE space for comparison.

\begin{figure*}[!th]
\centering
\begin{minipage}[t]{0.24\linewidth}
  \includegraphics[width=1.0\linewidth]{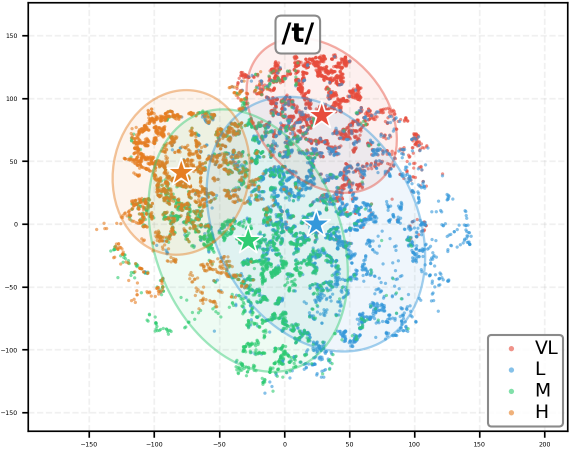}
\end{minipage}\hfill
\begin{minipage}[t]{0.24\linewidth}
  \includegraphics[width=1.0\linewidth]{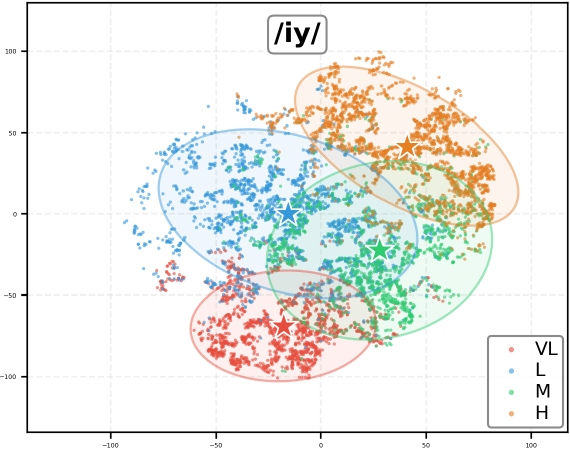}
\end{minipage}\hfill
\begin{minipage}[t]{0.24\linewidth}
  \includegraphics[width=1.0\linewidth]{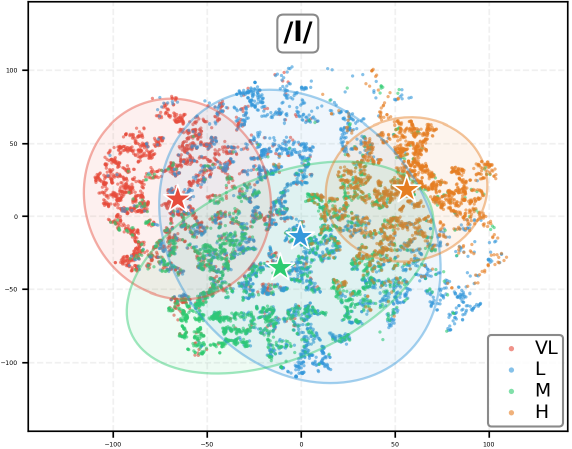}
\end{minipage}\hfill
\begin{minipage}[t]{0.24\linewidth}
  \includegraphics[width=1.0\linewidth]{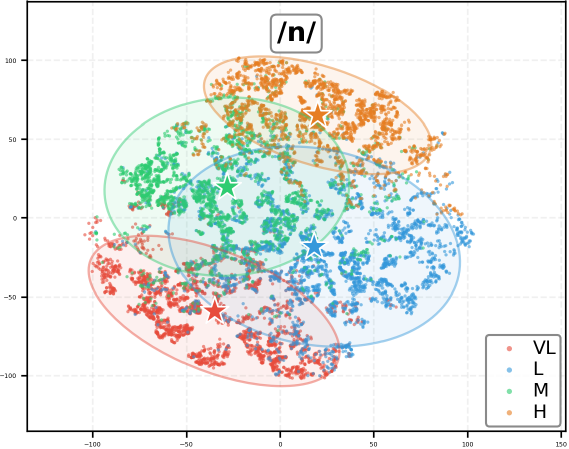}
\end{minipage}\\[1mm]
\begin{minipage}[t]{0.24\linewidth}
  \includegraphics[width=1.0\linewidth]{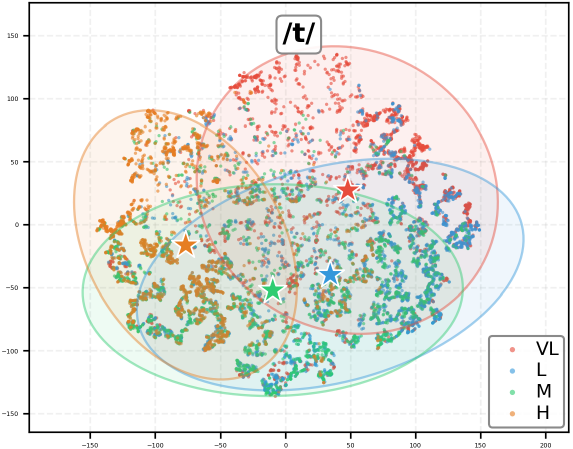}
\end{minipage}\hfill
\begin{minipage}[t]{0.24\linewidth}
  \includegraphics[width=1.0\linewidth]{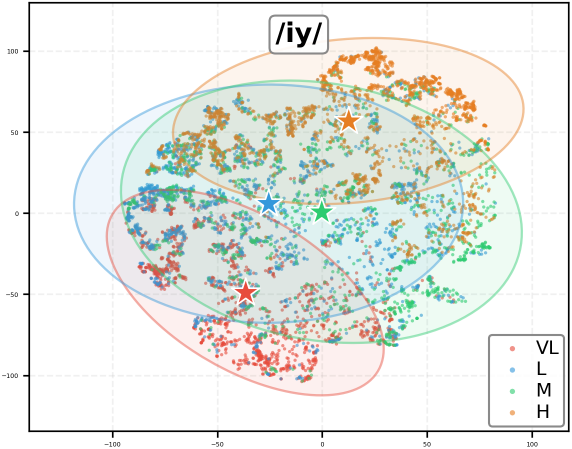}
\end{minipage}\hfill
\begin{minipage}[t]{0.24\linewidth}
  \includegraphics[width=1.0\linewidth]{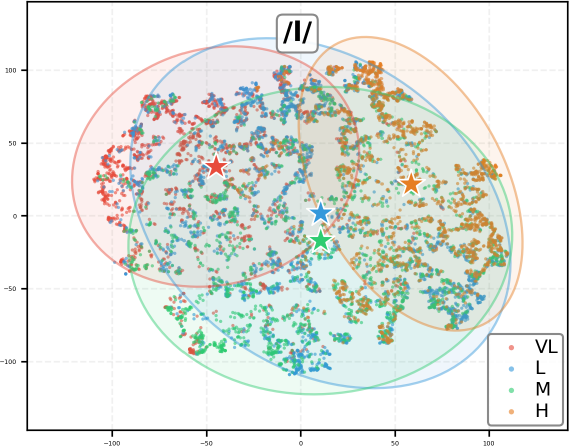}
\end{minipage}\hfill
\begin{minipage}[t]{0.24\linewidth}
  \includegraphics[width=1.0\linewidth]{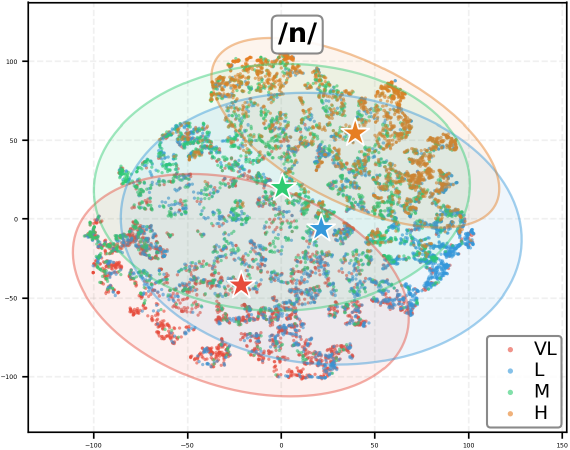}
\end{minipage}
\vspace{-4mm}
\caption{
\textbf{T-SNE visualization of phoneme-specific representations across dysarthric speech severity groups} for four representative phonemes (/t/, /iy/, /l/, and /n/).
\textbf{Top row:} without the proposed severity-invariant regularization; \textbf{Bottom row:} with the regularization.
Ellipses denote group-level confidence regions, and stars ($\bigstar$) mark group centroids.
}
\label{tsne}
\vspace{-4mm}
\end{figure*}

\begin{table}[!t]
\caption{
  \textbf{Efficiency comparison of speech representations on Conformer ASR.}
  SSL: continuous HuBERT features; Discrete: HuBERT K-means tokens ($K=100$);
  Dis.+Com.: discrete tokens with deduplication and BPE compression (BPE vocabulary size: 200).
  Rel. Len.: sequence length normalized to uncompressed SSL.
  Training time of model convergence is measured in GPU hours on a single NVIDIA A40 GPU. Inference RTF (Real Time Factor) is measured on CPU and reported separately on the UASpeech and TORGO test sets.
}
\vspace{-2mm}
\label{tab:efficiency}
\centering
\scriptsize
\setlength{\tabcolsep}{2.6pt}
\renewcommand{\arraystretch}{0.95}
\resizebox{0.85\columnwidth}{!}{%
\begin{tabular}{c|c|c|c|c|c}
\Xhline{1.0pt}
Dataset & Feature & Rel. Len. & Train (h) & RTF & WER (\%) \\
\Xhline{1.0pt}

\multirow{3}{*}{UASpeech}
& SSL
& 1.00$\times$
& 5.16
& 0.63
& 22.85 \\

& Discrete
& 1.00$\times$
& 2.67
& 0.60
& 24.29 \\

& Dis.+Com.
& 0.46$\times$
& 2.01
& 0.44
& 25.01 \\
\Xhline{1.0pt}

\multirow{3}{*}{TORGO}
& SSL
& 1.00$\times$
& 2.11
& 1.14
& 8.94 \\

& Discrete
& 1.00$\times$
& 1.78
& 1.13
& 9.66 \\

& Dis.+Com.
& 0.49$\times$
& 1.42
& 0.83
& 11.83 \\
\Xhline{1.0pt}
\end{tabular}
}
\vspace{-5mm}
\end{table}

\begin{table*}[!ht]
    \caption{
    \textbf{Performance of TDNN, Conformer (CONF.), Qwen2.5 based SpeechLLM systems, and system combination (Sys. Comb.) on UASpeech corpus.}
    ``HuBERT SSL'' denotes continuous HuBERT SSL representations, while ``HuBERT Discrete Token'' denotes discrete tokens extracted from HuBERT features using either conventional K-means or differentiable K-means (Diff. KM) quantization.
    ``K'' denotes codebook size for discrete token extraction.
    ``Token refinement'', ``E2E Optim.'', and ``Severity-invariant Reg.'' denote iterative pseudo-label update, differentiable end-to-end optimization, and severity-invariant regularization, respectively.
    ``VL'', ``L'', ``M'', and ``H'' denote the very-low, low, medium, and high intelligibility subgroups.
    ``$\dag$'' and ``$\star$'' indicate statistically significant improvements (MAPSSWE \cite{gillick1989some}, $\alpha = 0.05$) over the corresponding continuous SSL baseline (Sys. 1) and standard discrete token baselines (Sys. 3 and Sys. 8), respectively.
    For system combination, ``+'' denotes score interpolation, while ``X$\rightarrow$Y'' denotes two-pass rescoring \cite{cui2022two} of the $N$-best hypotheses ($N=100$) generated by system X using system Y.
    }
    \vspace{-2mm}
    \label{table:1}
    \centering
    \begin{tabular}{c|c|c|c|c|c|c|cccc|c}
    \Xhline{1.0pt}
    \multirow{3}{*}{Sys. \#} & 
    \multirow{3}{*}{Feature} &
    \multirow{3}{*}{\tabincell{c}{ASR\\Backend}} & 
    \multirow{3}{*}{Quantizer} &
    \multirow{3}{*}{\tabincell{c}{Token \\refinement}} &
    \multirow{3}{*}{\tabincell{c}{E2E\\Optim.}} &
    \multirow{3}{*}{\tabincell{c}{Severity-\\invariant\\Reg.}} & 
    \multicolumn{5}{c}{Word Error Rate (WER\%)} \\
    \cline{8-12}
     & & & & & & &
     \multicolumn{4}{c|}{Intelligibility Subgroup} &
     \multirow{2}{*}{Avg.} \\
     \cline{8-11}
     & & & & & & & H & M & L & VL & \\
     \Xhline{1.0pt}

    0-a & \multirow{2}{*}{FBank} & TDNN & \multirow{2}{*}{--} & \multicolumn{3}{c|}{\multirow{2}{*}{\xmark}} & 6.50 & 15.82 & 24.56 & 61.62 & 24.64 \\
    \cline{1-1}\cline{3-3}\cline{8-12}
    0-b & & CONF. & & \multicolumn{3}{c|}{} & 8.40 & 31.72 & 41.33 & 66.57 & 33.74 \\ 
    \Xhline{1.0pt}
    
    1 & \multirow{2}{*}{\tabincell{c}{HuBERT\\SSL}} & \multirow{2}{*}{CONF.} & \multirow{2}{*}{--} & \multirow{2}{*}{\xmark} & \xmark & \multirow{2}{*}{\xmark} & 3.15 & 14.01 & 24.86 & 59.73 & 22.85 \\
    \cline{6-6}
    2 & & & & & \cmark & & 3.24 & 12.21 & 22.57 & 55.09 & 20.96 \\
    \Xhline{1.0pt}
    
    3 & \multirow{5}{*}{\tabincell{c}{HuBERT\\Discrete\\Token\\($K=100$)}} & \multirow{10}{*}{CONF.} & K-means & \multicolumn{3}{c|}{\multirow{2}{*}{\xmark}} & 3.43 & 15.37 & 26.39 & 63.01 & 24.29 \\
    \cline{1-1}\cline{4-4}\cline{8-12}
    4 & & & \multirow{4}{*}{\tabincell{c}{Diff.\\KM}} & \multicolumn{3}{c|}{} & 3.28 & 15.13 & 26.47 & 62.95 & 24.21 \\
    \cline{5-7}
    
    5 & & & & \multirow{3}{*}{\cmark} & \xmark & \xmark & 3.24$^\star$ & 14.56$^\star$ & 26.62 & 62.24 & 23.97$^\star$ \\
    \cline{6-7}
    
    6 & & &  &  & \multirow{2}{*}{\cmark} & \xmark & 3.45 & 14.13$^\star$ & 24.69$^\star$ & 57.61$^{\star\dag}$ & 22.48$^{\star\dag}$  \\
    \cline{7-7}
    7 & & &  &  & & \cmark & 3.57 & 14.29$^\star$ & 24.25$^{\star\dag}$ & 55.90$^{\star\dag}$ & 22.07$^{\star\dag}$ \\
    \cline{1-2} \cline{4-12}

    8 & \multirow{7}{*}{\tabincell{c}{HuBERT\\Discrete\\Token\\($K=500$)}} & & K-means & \multicolumn{3}{c|}{\multirow{2}{*}{\xmark}} & 3.36 & 15.07 & 26.17 & 62.04 & 23.95 \\
    \cline{1-1}\cline{4-4}\cline{8-12}
    9 & & & \multirow{4}{*}{\tabincell{c}{Diff.\\KM}} & \multicolumn{3}{c|}{} & 3.27 & 15.19 & 26.40 & 61.90 & 23.97 \\
    \cline{5-7}
    
    10 & & & & \multirow{3}{*}{\cmark} & \xmark & \xmark &3.30 & 15.07 & 26.31 & 60.87$^\star$ & 23.72 \\
    \cline{6-7}
    
    11 & & &  &  & \multirow{2}{*}{\cmark} & \xmark & 3.24 & 14.35$^\star$ & 24.92$^\star$ & 56.98$^{\star\dag}$ & 22.38$^{\star\dag}$ \\
    \cline{7-7}
    12 & & &  &  & & \multirow{1}{*}{\cmark} &  3.41 & 14.37$^\star$ & 24.67$^\star$ & 55.88$^{\star\dag}$ & 22.09$^{\star\dag}$ \\
    \cline{1-1}\cline{3-12}
    
    13 & & \multirow{2}{*}{Qwen2.5} & K-means & \multicolumn{3}{c|}{\xmark} & 3.26 & 14.79 & 27.24 & 64.01 & 24.55 \\
    \cline{4-12}
    14 & & & Diff. KM & \multicolumn{3}{c|}{\cmark} & 3.66 & 14.78 & 25.40 & 56.23 & 22.57 \\
    \Xhline{1.0pt}

    15 & \multirow{2}{*}{Sys. Comb.} & \multicolumn{5}{c|}{Sys. 0-a + (0-a $\rightarrow$ 2)} & 3.17 & 8.45$^{\star\dag}$ & 18.11$^{\star\dag}$ & 50.82$^{\star\dag}$ & 18.14$^{\star\dag}$  \\
    16 & & \multicolumn{5}{c|}{Sys. 0-a + (0-a $\rightarrow$ 7) + (0-a $\rightarrow$ 12)} & 3.33 & 9.53$^{\star\dag}$ & 19.75$^{\star\dag}$ & 51.16$^{\star\dag}$ & 18.90$^{\star\dag}$  \\
    \Xhline{1.0pt}
    \end{tabular}
    \vspace{-5mm}
\end{table*}

\begin{table*}[!ht]
    \caption{
    \textbf{System performance on the TORGO test set.}
    ``$\dag$'' and ``$\star$'' indicate statistically significant improvements
    (MAPSSWE \cite{gillick1989some}, $\alpha = 0.05$) over the corresponding
    continuous SSL baseline (Sys. 1) and standard discrete token baseline (Sys. 3),
    respectively.
    Other naming conventions follow Table \ref{table:1}.
    }
    \vspace{-2mm}
    \label{table:2}
    \centering
    \setlength{\tabcolsep}{7pt}
    \begin{tabular}{c|c|c|c|c|c|c|ccc|c}
    \Xhline{1.0pt}
    \multirow{3}{*}{Sys. \#} &
    \multirow{3}{*}{Feature} &
    \multirow{3}{*}{\tabincell{c}{ASR\\Backend}} &
    \multirow{3}{*}{Quantizer} &
    \multirow{3}{*}{\tabincell{c}{Token\\refinement}} &
    \multirow{3}{*}{\tabincell{c}{E2E\\Optim.}} &
    \multirow{3}{*}{\tabincell{c}{Severity-\\invariant\\Reg.}} &
    \multicolumn{4}{c}{Word Error Rate (WER\%)} \\
    \cline{8-11}
    & & & & & & &
    \multicolumn{3}{c|}{Intelligibility Subgroup} &
    \multirow{2}{*}{Avg.} \\
    \cline{8-10}
    & & & & & & & Mild & Moderate & Severe & \\
    \Xhline{1.0pt}

    0-a & \multirow{2}{*}{FBank} &
    TDNN &
    \multirow{2}{*}{--} &
    \multicolumn{3}{c|}{\multirow{2}{*}{\xmark}} &
    3.25 & 8.27 & 12.52 & 9.11 \\
    \cline{1-1}\cline{3-3}\cline{8-11}

    0-b & &
    CONF. &
    & \multicolumn{3}{c|}{} &
    4.80 & 6.63 & 21.22 & 13.72 \\
    \Xhline{1.0pt}

    1 & \multirow{2}{*}{\tabincell{c}{HuBERT\\SSL}} &
    \multirow{2}{*}{CONF.} &
    \multirow{2}{*}{--} &
    \multirow{2}{*}{\xmark} &
    \xmark &
    \multirow{2}{*}{\xmark} &
    2.86 & 4.08 & 14.07 & 8.94 \\
    \cline{6-6}

    2 & & & & &
    \cmark &
    &
    2.63 & 4.18 & 11.18 & 7.40 \\
    \Xhline{1.0pt}

    3 & \multirow{7}{*}{\tabincell{c}{HuBERT\\Discrete\\Token\\($K=100$)}} &
    \multirow{5}{*}{CONF.} &
    K-means &
    \multicolumn{3}{c|}{\multirow{2}{*}{\xmark}} &
    2.63 & 4.29 & 15.49 & 9.66 \\
    \cline{1-1}\cline{4-4}\cline{8-11}

    4 & & &
    \multirow{4}{*}{\tabincell{c}{Diff.\\KM}} &
    \multicolumn{3}{c|}{} &
    2.40 & 4.39 & 15.61 & 9.68 \\
    \cline{5-7}

    5 & & & &
    \multirow{3}{*}{\cmark} &
    \xmark &
    \xmark &
    2.55 & 4.59 & 15.28 & 9.59 \\
    \cline{6-7}

    6 & & & &
    &
    \multirow{2}{*}{\cmark} &
    \xmark &
    2.85 & 4.18 & 12.32$^{\star\dag}$ & 8.05$^{\star\dag}$ \\
    \cline{7-7}

    7 & & & &
    &
    &
    \cmark &
    2.63 & 4.18 & 12.11$^{\star\dag}$ & 7.88$^{\star\dag}$ \\

    \cline{1-1}\cline{3-11}
    
    8 & &
    \multirow{2}{*}{Qwen2.5} &
    K-means &
    \multicolumn{3}{c|}{\xmark} &
    2.40 & 5.71 & 16.07 & 10.21 \\
    \cline{4-11}

    9 & & &
    Diff. KM &
    \multicolumn{3}{c|}{\cmark} &
    2.64 & 5.81 & 13.41 & 8.74 \\
    \Xhline{1.0pt}

    10 & \multirow{2}{*}{Sys. Comb.} &
    \multicolumn{5}{c|}{Sys. 0-a + (0-a $\rightarrow$ 2)} &
    2.63 & 3.78$^{\star\dag}$ & 8.98$^{\star\dag}$ & 6.17$^{\star\dag}$ \\

    11 & &
    \multicolumn{5}{c|}{Sys. 0-a + (0-a $\rightarrow$ 7)} &
    2.79 & 3.67$^{\star\dag}$ & 9.35$^{\star\dag}$ & 6.38$^{\star\dag}$ \\
    
    \Xhline{1.0pt}
    \end{tabular}
    \vspace{-4mm}
\end{table*}

\begin{table*}[!ht]
    \centering
    \caption{\textbf{Performance comparison of published systems on UASpeech and ours.} Naming conventions follow Table \ref{table:1}.
    }
    \vspace{-2mm}
    \label{table:3}
    \begin{tabular}{c|c|c|c}
    \Xhline{1pt}
    Performance (WER\%) of Systems Published on UASpeech& L & VL & All \\
    \Xhline{1pt}
    CUHK-2021 DNN + DCGAN + LHUC-SAT \cite{jin21_interspeech} & 27.37 & 61.42 & 25.89 \\
    Brno Univ.-2022 Wav2vec2 + SAT (15 spkr) \cite{baskar22b_interspeech} & 22.46 & 57.72 & 22.83\\
    FAU-2022 Cross-lingual XLRS + Conformer \cite{hernandez22_interspeech} & 28.60 & 62.00 & 26.10 \\
    CUHK-2023 Kaldi TDNN + VAE-GAN + LHUC-SAT \cite{jin23vae} & 28.53 & 57.31 & 27.78 \\
    JHU-2023 DuTa-VC (Diffusion) + Conformer \cite{wang23qa_interspeech} & 27.70 & 63.70 & 27.90\\
    CUHK-2023 TDNN + Wav2vec2.0 + Sys. Comb. \cite{hu2023exploring} & 25.03 & 53.12 & 22.56 \\
    CUHK-2023 DNN + Wav2vec2.0 + Sys. Comb. + Sev. Adapt \cite{geng23b_interspeech} & 17.41 & 51.25 & 17.82 \\
    CUHK- 2025 TDNN/Conformer + PPG Discrete Token + Sys. Comb.  \cite{huimeng2025icassp}& 25.68 & 54.14 & 23.25 \\
    NCKU- 2025 Conformer + Curriculum Learning + Multi-stream feature fusion \cite{hsieh25_interspeech} & 24.55  & 55.96 & 21.55 \\
    \hline
    \textbf{Ours, TDNN/Conformer + Seve.-Invar. Discrete Token + Sys. Comb.} & \textbf{19.75} & \textbf{51.16} & \textbf{18.90} \\
    \Xhline{1pt}
\end{tabular}
\vspace{-5mm}
\end{table*}

\subsection{Result Analysis}
\label{subsection:result}
Table~\ref{tab:efficiency} presents an efficiency comparison across three speech representation configurations.
\textbf{1)} Compared to continuous SSL features, uncompressed discrete tokens reduce training time by 48\% on UASpeech (5.16h vs. 2.67h) and 16\% on TORGO (2.11h vs. 1.78h), with slightly faster inference RTF, and WER increases of 1.44\% and 0.72\%, respectively. 
\textbf{2)} Applying deduplication and BPE compression further reduces sequence length to 0.46/0.49$\times$ and training time by an additional 25\%/20\%, and also yields lower RTF (0.60 vs. 0.44 on UASpeech, 1.13 vs. 0.83 on TORGO), but at the cost of more pronounced WER degradation of 0.72\% and 2.17\%.
Given that recognition accuracy is the primary concern in our work, all subsequent experiments adopt uncompressed discrete tokens, accepting a moderate efficiency trade-off in favor of performance.
  
Results on the \textbf{UASpeech} test set (Table \ref{table:1}) show that:
\textbf{1)}
\textbf{Vs. Standard Discrete Tokens:} The proposed DSI tokens consistently and significantly outperform Conformer systems using standard K-means tokens across varying codebook sizes, achieving up to a \textbf{2.22\% absolute (9.14\% relative)} WER reduction (Sys. 7 vs. 3). The gains are particularly pronounced in the most challenging VL subgroup, where WER decreases from 63.01\% to 55.90\% (\textbf{7.11\% absolute}).
\textbf{2) Vs. Continuous SSL Features:} The DSI discrete tokens also significantly surpass the continuous HuBERT SSL baseline on Conformer systems, yielding up to a \textbf{0.78\% absolute (3.41\% relative)} WER reduction (Sys. 7 vs. 1).
This advantage is again concentrated in most severe speakers, with VL WER improving from 59.73\% to 55.90\% (\textbf{3.83\% absolute}).
\textbf{3) Impact of Proposed Methods:}
Pseudo-label update based token refinement consistently improves the initial DKM, reducing WERs to \textbf{23.97\%} (Sys. 5, $K=100$) and \textbf{23.72\%} (Sys. 10, $K=500$).
E2E optimization substantially improves the non-E2E counterparts by \textbf{1.49\%} (Sys. 6 vs. 5, $K=100$) and \textbf{1.34\%} (Sys. 11 vs. 10, $K=500$) in absolute WER.
The SIR method yields \textbf{0.41\%}/\textbf{0.29\%} absolute WER reductions for $K=100/500$ (Sys. 7 vs. 6 / Sys. 12 vs. 11), with gains predominantly concentrated in the most severe VL subgroup (\textbf{1.71\%}/\textbf{1.10\%} absolute), while leaving higher-intelligibility subgroups largely unaffected.
\textbf{4)} \textbf{The effectiveness of SIR} is further confirmed by phoneme-specific T-SNE visualizations (Fig. \ref{tsne}), where severity-group distributions exhibit greater overlap and less distinct boundaries after its application, suggesting reduced severity-dependent variation.
\textbf{5)} The DSI framework benefits the Qwen-based SpeechLLM backend, yielding a \textbf{1.98\% absolute} WER reduction over K-means token baseline (Sys.~14 vs.~13).
\textbf{6)} Fusing the DSI token systems with the FBank baseline produces our lowest WER of \textbf{18.90\%} (Sys. 16), which closely matches its continuous SSL counterpart (Sys. 15, 18.14\%) and recently published continuous systems (Table \ref{table:3}).
Notably, our system achieves a \textbf{VL subgroup WER of 51.16\%}, outperforming most systems in Table~\ref{table:3} on this most challenging subgroup, while relying on discrete tokens.

Results on \textbf{TORGO} (Table \ref{table:2}) exhibit trends consistent with those on UASpeech. DSI discrete token system (Sys. 7) achieves a WER of \textbf{7.88\%}, yielding statistically significant absolute reductions of \textbf{1.78\%} and \textbf{1.06\%} over the standard K-means (Sys. 3) and continuous SSL (Sys. 1) baselines, respectively.
The improvements are predominantly concentrated in the Severe subgroup, with WER decreasing from 15.49\%/14.07\% (Sys. 3/ Sys. 1) to 12.11\% (\textbf{3.38\%/1.96\% absolute}), while Mild and Moderate speakers show marginal changes. Token refinement provides only modest gains (Sys. 5 vs. 4), whereas E2E optimization drives the largest improvement, reducing Severe WER by \textbf{2.96\% absolute} (Sys. 6 vs. 5, 12.32\% vs. 15.28\%). SIR contributes a further \textbf{0.21\%} absolute reduction in the Severe subgroup (Sys. 7 vs. 6, 12.11\% vs. 12.32\%).
On Qwen-based SpeechLLM systems, DSI tokens also reduce WER by \textbf{1.47\%} absolute over the K-means token baseline (Sys. 9 vs. 8).
System combination (Sys.11) produces the lowest WER of \textbf{6.38\%} on the test set, closely matching its continuous SSL counterpart (Sys. 10, 6.17\%).

\section{Conclusion}
\label{sec:conclusion}
\vspace{-1mm}
This paper proposed differentiable and severity-invariant discrete tokens for dysarthric speech recognition. The proposed framework combines iterative pseudo-label refinement, end-to-end optimization, and severity-invariant regularization to learn ASR-oriented representations with reduced severity-dependent variation. Experiments on UASpeech and TORGO demonstrate consistent improvements over conventional K-means tokens and comparable continuous HuBERT SSL features. Future work will explore finer-grained local pathological speech characteristics for representation learning.

\section*{Acknowledgement}
\vspace{-1mm}
This research is supported by Hong Kong RGC GRF grant No. 
14200021, 14200324, Innovation Technology Fund grant No. ITS/218/21, Basic Research Project of Institute of Software, Chinese Academy of Sciences ISCAS-JCMS-202306, and Youth Innovation Promotion Association CAS Grant 2023119.

\newpage

\bibliographystyle{IEEEtran}
\bibliography{reference}

@INPROCEEDINGS{xiong20source,
  author={Xiong, Feifei and Barker, Jon and Yue, Zhengjun and Christensen, Heidi},
  booktitle={ICASSP 2020 - 2020 IEEE International Conference on Acoustics, Speech and Signal Processing (ICASSP)}, 
  title={Source Domain Data Selection for Improved Transfer Learning Targeting Dysarthric Speech Recognition}, 
  year={2020},
  volume={},
  number={},
}

@article{mengzhetaslp,
author = {Geng, Mengzhe and Xie, Xurong and Ye, Zi and Wang, Tianzi and Li, Guinan and Hu, Shujie and Liu, Xunying and Meng, Helen},
title = {Speaker Adaptation Using Spectro-Temporal Deep Features for Dysarthric and Elderly Speech Recognition},
year = {2022},
issue_date = {2022},
publisher = {IEEE Press},
volume = {30},
journal = {IEEE/ACM Trans. Audio, Speech and Lang. Proc.},
month = jul,
pages = {2597–2611},
numpages = {15}
}

@article{shujie23taslp,
author = {Hu, Shujie and Xie, Xurong and Geng, Mengzhe and Jin, Zengrui and Deng, Jiajun and Li, Guinan and Wang, Yi and Cui, Mingyu and Wang, Tianzi and Meng, Helen and Liu, Xunying},
title = {Self-Supervised ASR Models and Features for Dysarthric and Elderly Speech Recognition},
year = {2024},
issue_date = {2024},
publisher = {IEEE Press},
volume = {32},
journal = {IEEE/ACM Trans. Audio, Speech and Lang. Proc.},
month = jul,
pages = {3561–3575},
numpages = {15}
}

@article{wang2023benefits,
  author    = {Wang, Pu and Hugo Van hamme},
  title     = {{Benefits of Pre-Trained Mono- and Cross-Lingual Speech Representations for Spoken Language Understanding of Dutch Dysarthric Speech}},
  journal   = {EURASIP J. Audio Speech Music Process.},
  year      = {2023}
}

@INPROCEEDINGS{huimeng2024icassp,
  author={Wang, Huimeng and Jin, Zengrui and Geng, Mengzhe and Hu, Shujie and Li, Guinan and Wang, Tianzi and Xu, Haoning and Liu, Xunying},
  booktitle={ICASSP 2024 - 2024 IEEE International Conference on Acoustics, Speech and Signal Processing (ICASSP)}, 
  title={Enhancing Pre-Trained ASR System Fine-Tuning for Dysarthric Speech Recognition Using Adversarial Data Augmentation}, 
  year={2024},
  volume={},
  number={},
}

@INPROCEEDINGS{huimeng2025icassp,
  author={Wang, Huimeng and Xie, Xurong and Geng, Mengzhe and Hu, Shujie and Xu, Haoning and Chen, Youjun and Li, Zhaoqing and Deng, Jiajun and Liu, Xunying},
  booktitle={ICASSP 2025 - 2025 IEEE International Conference on Acoustics, Speech and Signal Processing (ICASSP)}, 
  title={Phone-purity Guided Discrete Tokens for Dysarthric Speech Recognition}, 
  year={2025},
  volume={},
  number={},
}

@inproceedings{hu25d_interspeech,
  title     = {{On-the-fly Routing for Zero-shot MoE Speaker Adaptation of Speech Foundation Models for Dysarthric Speech Recognition}},
  author    = {Shujie Hu and Xurong Xie and Mengzhe Geng and Jiajun Deng and Huimeng Wang and Guinan Li and Chengxi Deng and Tianzi Wang and Mingyu Cui and Helen Meng and Xunying Liu},
  year      = {2025},
  booktitle = {{Interspeech 2025}},
}

@inproceedings{aboeitta25_interspeech,
  title     = {{Bridging ASR and LLMs for Dysarthric Speech Recognition: Benchmarking Self-Supervised and Generative Approaches }},
  author    = {Ahmed Aboeitta and Ahmed Sharshar and Youssef Nafea and Shady Shehata},
  year      = {2025},
  booktitle = {{Interspeech 2025}},
}

@inproceedings{xiong25_interspeech,
  title     = {{Mitigating Overfitting During Speech Foundation Model Fine-tuning: Applications to Dysarthric Speech Detection}},
  author    = {Yan Xiong and Visar Berisha and Julie Liss and Chaitali Chakrabarti},
  year      = {2025},
  booktitle = {{Interspeech 2025}},
}

@inproceedings{hsieh25_interspeech,
  title     = {{Dysarthric Speech Recognition Using Curriculum Learning and Multi-stream Architecture}},
  author    = {I-Ting Hsieh and Chung-Hsien Wu},
  year      = {2025},
  booktitle = {{Interspeech 2025}},
}

@inproceedings{baevski2020wav2vec,
author = {Baevski, Alexei and Zhou, Henry and Mohamed, Abdelrahman and Auli, Michael},
title = {wav2vec 2.0: a framework for self-supervised learning of speech representations},
year = {2020},
booktitle = {Proceedings of the 34th International Conference on Neural Information Processing Systems},
}

@ARTICLE{hsu2021hubert,
  author={Hsu, Wei-Ning and Bolte, Benjamin and Tsai, Yao-Hung Hubert and Lakhotia, Kushal and Salakhutdinov, Ruslan and Mohamed, Abdelrahman},
  journal={IEEE/ACM Transactions on Audio, Speech, and Language Processing}, 
  title={HuBERT: Self-Supervised Speech Representation Learning by Masked Prediction of Hidden Units}, 
  year={2021},
  volume={29},
  number={},
  pages={3451-3460},
}

@ARTICLE{chen2022wavlm,
  author={Chen, Sanyuan and Wang, Chengyi and Chen, Zhengyang and Wu, Yu and Liu, Shujie and Chen, Zhuo and Li, Jinyu and Kanda, Naoyuki and Yoshioka, Takuya and Xiao, Xiong and Wu, Jian and Zhou, Long and Ren, Shuo and Qian, Yanmin and Qian, Yao and Wu, Jian and Zeng, Michael and Yu, Xiangzhan and Wei, Furu},
  journal={IEEE Journal of Selected Topics in Signal Processing}, 
  title={WavLM: Large-Scale Self-Supervised Pre-Training for Full Stack Speech Processing}, 
  year={2022},
  volume={16},
  number={6},
  pages={1505-1518},
}

@inproceedings{chang23b_interspeech,
  title     = {{Exploration of Efficient End-to-End ASR using Discretized Input from Self-Supervised Learning}},
  author    = {Xuankai Chang and Brian Yan and Yuya Fujita and Takashi Maekaku and Shinji Watanabe},
  year      = {2023},
  booktitle = {{Interspeech 2023}},
}

@INPROCEEDINGS{chang2024exploring,
  author={Chang, Xuankai and Yan, Brian and Choi, Kwanghee and Jung, Jee-Weon and Lu, Yichen and Maiti, Soumi and Sharma, Roshan and Shi, Jiatong and Tian, Jinchuan and Watanabe, Shinji and Fujita, Yuya and Maekaku, Takashi and Guo, Pengcheng and Cheng, Yao-Fei and Denisov, Pavel and Saijo, Kohei and Wang, Hsiu-Hsuan},
  booktitle={ICASSP 2024 - 2024 IEEE International Conference on Acoustics, Speech and Signal Processing (ICASSP)}, 
  title={Exploring Speech Recognition, Translation, and Understanding with Discrete Speech Units: A Comparative Study}, 
  year={2024},
  volume={},
  number={},
}

@INPROCEEDINGS{yang2024towards,
  author={Yang, Yifan and Shen, Feiyu and Du, Chenpeng and Ma, Ziyang and Yu, Kai and Povey, Daniel and Chen, Xie},
  booktitle={ICASSP 2024 - 2024 IEEE International Conference on Acoustics, Speech and Signal Processing (ICASSP)}, 
  title={Towards Universal Speech Discrete Tokens: A Case Study for ASR and TTS}, 
  year={2024},
  volume={},
  number={},
}

@inproceedings{onda25c_interspeech,
  title     = {{Differentiable K-means for Fully-optimized Discrete Token-based ASR}},
  author    = {Kentaro Onda and Yosuke Kashiwagi and Emiru Tsunoo and Hayato Futami and Shinji Watanabe},
  year      = {2025},
  booktitle = {{Interspeech 2025}},
}

@inproceedings{du22b_interspeech,
  title     = {{VQTTS: High-Fidelity Text-to-Speech Synthesis with Self-Supervised VQ Acoustic Feature}},
  author    = {Chenpeng Du and Yiwei Guo and Xie Chen and Kai Yu},
  year      = {2022},
  booktitle = {{Interspeech 2022}},
}

@inproceedings{unicats,
author = {Du, Chenpeng and Guo, Yiwei and Shen, Feiyu and Liu, Zhijun and Liang, Zheng and Chen, Xie and Wang, Shuai and Zhang, Hui and Yu, Kai},
title = {UniCATS: a unified context-aware text-to-speech framework with contextual VQ-diffusion and vocoding},
year = {2024},
booktitle = {Proceedings of the Thirty-Eighth AAAI Conference on Artificial Intelligence and Thirty-Sixth Conference on Innovative Applications of Artificial Intelligence and Fourteenth Symposium on Educational Advances in Artificial Intelligence},
}

@inproceedings{wang25ba_interspeech,
  title     = {{Discl-VC: Disentangled Discrete Tokens and In-Context Learning for Controllable Zero-Shot Voice Conversion}},
  author    = {Kaidi Wang and Wenhao Guan and Ziyue Jiang and Hukai Huang and Peijie Chen and Weijie Wu and Qingyang Hong and Lin Li},
  year      = {2025},
  booktitle = {{Interspeech 2025}},
}

@inproceedings{bai25_interspeech,
  title     = {{Accent Normalization Using Self-Supervised Discrete Tokens with Non-Parallel Data}},
  author    = {Qibing Bai and Sho Inoue and Shuai Wang and Zhongjie Jiang and Yannan Wang and Haizhou Li},
  year      = {2025},
  booktitle = {{Interspeech 2025}},
}

@inproceedings{sukhadia24_interspeech,
  title     = {{Children’s Speech Recognition through Discrete Token Enhancement}},
  author    = {Vrunda N. Sukhadia and Shammur Absar Chowdhury},
  year      = {2024},
  booktitle = {{Interspeech 2024}},
}

@inproceedings{yeh2024slt,
  author    = {Yeh, Sung-Lin and Tang, Hao},
  title     = {Estimating the Completeness of Discrete Speech Units},
  booktitle = {Proc. SLT},
  year      = {2024}
}

@inproceedings{gulati20_interspeech,
  title     = {{Conformer: Convolution-augmented Transformer for Speech Recognition}},
  author    = {Anmol Gulati and James Qin and Chung-Cheng Chiu and Niki Parmar and Yu Zhang and Jiahui Yu and Wei Han and Shibo Wang and Zhengdong Zhang and Yonghui Wu and Ruoming Pang},
  year      = {2020},
  booktitle = {{Interspeech 2020}},
}

@inproceedings{uaspeech2008,
  title     = {{Dysarthric speech database for universal access research}},
  author    = {Heejin Kim and Mark Hasegawa-Johnson and Adrienne Perlman and Jon Gunderson and Thomas S. Huang and Kenneth Watkin and Simone Frame},
  year      = {2008},
  booktitle = {{Interspeech 2008}},
}

@article{torgo,
  author    = {Rudzicz, Frank and Namasivayam, Aravind Kumar and Wolff, Talya},
  title     = {The TORGO database of acoustic and articulatory speech from speakers with dysarthria},
  journal   = {Lang. Resour. Eval.},
  year      = {2012}
}

@inproceedings{jang2016categorical,
  author    = {Jang, Eric and Gu, Shixiang and Poole, Ben},
  title     = {Categorical Reparameterization with Gumbel-Softmax},
  booktitle = {Proc. ICLR},
  year      = {2017}
}

@INPROCEEDINGS{mengistu2011adapting,
  author={Mengistu, Kinfe Tadesse and Rudzicz, Frank},
  booktitle={2011 IEEE International Conference on Acoustics, Speech and Signal Processing (ICASSP)}, 
  title={Adapting acoustic and lexical models to dysarthric speech}, 
  year={2011},
  volume={},
  number={},
}

@INPROCEEDINGS{gao2020deep,
  author={Gao, Boyan and Yang, Yongxin and Gouk, Henry and Hospedales, Timothy M.},
  booktitle={ICASSP 2020 - 2020 IEEE International Conference on Acoustics, Speech and Signal Processing (ICASSP)}, 
  title={Deep Clusteringwith Concrete K-Means}, 
  year={2020},
  volume={},
  number={},
}

@inproceedings{hernandez22_interspeech,
  title     = {{Cross-lingual Self-Supervised Speech Representations for Improved Dysarthric Speech Recognition}},
  author    = {Abner Hernandez and Paula Andrea Pérez-Toro and Elmar Noeth and Juan Rafael Orozco-Arroyave and Andreas Maier and Seung Hee Yang},
  year      = {2022},
  booktitle = {{Interspeech 2022}},
}

@ARTICLE{10081405,
  author={Yu, Chongchong and Su, Xiaosu and Qian, Zhaopeng},
  journal={IEEE Transactions on Neural Systems and Rehabilitation Engineering}, 
  title={Multi-Stage Audio-Visual Fusion for Dysarthric Speech Recognition With Pre-Trained Models}, 
  year={2023},
  volume={31},
  number={},
  pages={1912-1921},
}

@inproceedings{jin21_interspeech,
  title     = {{Adversarial Data Augmentation for Disordered Speech Recognition}},
  author    = {Zengrui Jin and Mengzhe Geng and Xurong Xie and Jianwei Yu and Shansong Liu and Xunying Liu and Helen Meng},
  year      = {2021},
  booktitle = {{Interspeech 2021}},
}

@inproceedings{baskar22b_interspeech,
  title     = {{Speaker adaptation for Wav2vec2 based dysarthric ASR}},
  author    = {Murali Karthick Baskar and Tim Herzig and Diana Nguyen and Mireia Diez and Tim Polzehl and Lukas Burget and Jan Černocký},
  year      = {2022},
  booktitle = {{Interspeech 2022}},
}

@INPROCEEDINGS{jin23vae,
  author={Jin, Zengrui and Xie, Xurong and Geng, Mengzhe and Wang, Tianzi and Hu, Shujie and Deng, Jiajun and Li, Guinan and Liu, Xunying},
  booktitle={ICASSP 2023 - 2023 IEEE International Conference on Acoustics, Speech and Signal Processing (ICASSP)}, 
  title={Adversarial Data Augmentation Using VAE-GAN for Disordered Speech Recognition}, 
  year={2023},
  volume={},
  number={},
}

@inproceedings{wang23qa_interspeech,
  title     = {{DuTa-VC: A Duration-aware Typical-to-atypical Voice Conversion Approach with Diffusion Probabilistic Model}},
  author    = {Helin Wang and Thomas Thebaud and Jesús Villalba and Myra Sydnor and Becky Lammers and Najim Dehak and Laureano Moro-Velazquez},
  year      = {2023},
  booktitle = {{Interspeech 2023}},
}

@INPROCEEDINGS{hu2023exploring,
  author={Hu, Shujie and Xie, Xurong and Jin, Zengrui and Geng, Mengzhe and Wang, Yi and Cui, Mingyu and Deng, Jiajun and Liu, Xunying and Meng, Helen},
  booktitle={ICASSP 2023 - 2023 IEEE International Conference on Acoustics, Speech and Signal Processing (ICASSP)}, 
  title={Exploring Self-Supervised Pre-Trained ASR Models for Dysarthric and Elderly Speech Recognition}, 
  year={2023},
  volume={},
  number={},
}

@inproceedings{geng23b_interspeech,
  title     = {{Use of Speech Impairment Severity for Dysarthric Speech Recognition}},
  author    = {Mengzhe Geng and Zengrui Jin and Tianzi Wang and Shujie Hu and Jiajun Deng and Mingyu Cui and Guinan Li and Jianwei Yu and Xurong Xie and Xunying Liu},
  year      = {2023},
  booktitle = {{Interspeech 2023}},
}

@article{young2002htk,
  author    = {Young, Steve and others},
  title     = {{The HTK Book}},
  journal   = {Cambridge University Engineering Department},
  year      = {2002}
}

@inproceedings{geng2020investigation,
  title     = {{Investigation of Data Augmentation Techniques for Disordered Speech Recognition}},
  author    = {Mengzhe Geng and Xurong Xie and Shansong Liu and Jianwei Yu and Shoukang Hu and Xunying Liu and Helen Meng},
  year      = {2020},
  booktitle = {{Interspeech 2020}},
}

@inproceedings{watanabe18_interspeech,
  title     = {{ESPnet: End-to-End Speech Processing Toolkit}},
  author    = {Shinji Watanabe and Takaaki Hori and Shigeki Karita and Tomoki Hayashi and Jiro Nishitoba and Yuya Unno and Nelson {Enrique Yalta Soplin} and Jahn Heymann and Matthew Wiesner and Nanxin Chen and Adithya Renduchintala and Tsubasa Ochiai},
  year      = {2018},
  booktitle = {{Interspeech 2018}},
}

@INPROCEEDINGS{gillick1989some,
  author={Gillick, L. and Cox, S.J.},
  booktitle={International Conference on Acoustics, Speech, and Signal Processing,}, 
  title={Some statistical issues in the comparison of speech recognition algorithms}, 
  year={1989},
  volume={},
  number={},
}

@inproceedings{cui2022two,
  title     = {{Two-pass Decoding and Cross-adaptation Based System Combination of End-to-end Conformer and Hybrid TDNN ASR Systems}},
  author    = {Mingyu Cui and Jiajun Deng and Shoukang Hu and Xurong Xie and Tianzi Wang and Shujie Hu and Mengzhe Geng and Boyang Xue and Xunying Liu and Helen Meng},
  year      = {2022},
  booktitle = {{Interspeech 2022}},
}

@inproceedings{povey16_interspeech,
  title     = {{Purely Sequence-Trained Neural Networks for ASR Based on Lattice-Free MMI}},
  author    = {Daniel Povey and Vijayaditya Peddinti and Daniel Galvez and Pegah Ghahremani and Vimal Manohar and Xingyu Na and Yiming Wang and Sanjeev Khudanpur},
  year      = {2016},
  booktitle = {{Interspeech 2016}},
}

@inproceedings{cui25_interspeech,
  title     = {{Exploring SSL Discrete Speech Features for Zipformer-based Contextual ASR}},
  author    = {Mingyu Cui and Yifan Yang and Jiajun Deng and Jiawen Kang and Shujie Hu and Tianzi Wang and Zhaoqing Li and Shiliang Zhang and Xie Chen and Xunying Liu},
  year      = {2025},
  booktitle = {{Interspeech 2025}},
}

\end{document}